\documentclass{aa}  
\usepackage{graphicx}
  \usepackage{amsthm}
\usepackage{txfonts}
\usepackage{natbib}
\usepackage{xcolor}
\usepackage{url}
\usepackage{pifont}
\usepackage{comment}
\usepackage{amsthm}
\usepackage[colorlinks=true,
    linkcolor=blue,
    citecolor=blue,
    filecolor=magenta,      
    urlcolor=cyan]{hyperref}
\usepackage{subfigure}
\usepackage{subcaption}
\usepackage{amsmath}

\newcommand{\msun}{\mathrm{M_{\odot}}}

\newcommand{\lsim}{\mathrel{\rlap{\lower 3pt \hbox{$\sim$}} \raise 2.0pt \hbox{$<$}}}
\newcommand{\gsim}{\mathrel{\rlap{\lower 3pt \hbox{$\sim$}} \raise 2.0pt \hbox{$>$}}}

\begin{document}

   \title{Searching for unresolved massive black hole pairs through AGN photometric variability}

   \subtitle{}

   \author{Lorenzo Bertassi
          \inst{1,2,3}\fnmsep\thanks{l.bertassi@campus.unimib.it}
          \and 
          Maria Charisi \inst{5,6} 
          \and
          Fabio Rigamonti \inst{2,3,4} 
          \and
          Stefano Covino \inst{2,4}
          \and
          Massimo Dotti \inst{1,2,3}
          }
   \institute{Università degli Studi di Milano-Bicocca, Piazza della Scienza 3, 20126 Milano, Italy
    \and
    INAF - Osservatorio Astronomico di Brera, via Brera 20, I-20121 Milano, Italy
    \and
    INFN, Sezione di Milano-Bicocca, Piazza della Scienza 3, I-20126 Milano, Italy
    \and
    Como Lake centre for AstroPhysics (CLAP), DiSAT, Università dell’Insubria, via Valleggio 11, 22100 Como, Italy
    \and
    Department of Physics and Astronomy, Washington State University, Pullman, WA 99163, USA
    \and
    Institute of Astrophysics, FORTH, GR-71110, Heraklion, Greece
    }

   \date{Received ; accepted }
 
  \abstract{ 
  Since their discovery, active galactic nuclei light curves are known to be intrinsically variable. In the optical/UV band, this variability is consistent with correlated or red noise and is particularly well described by the damped random walk model. In this work, we evaluate the feasibility of a new method for identifying spatially unresolved couples of active galactic nuclei through a fully Bayesian time-domain analysis of the observed light curves. More specifically, we check whether observed light curves are better described by a single DRW, which we interpret as emitted by a single massive black hole, or a pair of independent damped random walks, generated by a pair of massive black holes. We test the method on mock light curves associated with single massive black holes and pairs generated with different cadences and lengths of observational campaigns. 
  We constrained the occurrence of false positives, that is, the percentage of single MBH light curves that show substantial evidence in favour of the unresolved MBH pair scenario, finding a fraction of $0.2\%$ and $0.59\%$ in the even and uneven sampling scenarios.
  We discuss how well the method recovers the model parameters, showing that about $51\%$ and $7 \%$ of the simulated light curves have all the recovered parameters within $20\%$ of their true values in our best scenario of evenly sampled light curves for the single massive black hole and massive black hole pair scenarios, respectively.  We finally study the region of the parameter space in which the detection of a massive black hole pair is possible, finding that such objects can be correctly identified if the timescales of the process describing the noise are very different, with a ratio smaller than $\sim 0.2$, and the variability amplitudes are similar, with their ratio bigger than $\sim 0.2$. When limiting to such a region of the parameter space, the fraction of pairs with all the recovered parameters within $20\%$ of the injected values increases up to 
  about $14\%$ and $8 \%$ for evenly and unevenly sampled light curves, respectively. 
  }


   \maketitle

\section{Introduction} \label{sec:introduction}
In the hierarchical model of galaxy formation, smaller galaxies merge to form the larger present-day galaxies \citep[][]{white1978}. As nearly all massive galaxies are expected to host a massive black hole (MBH) at their centre \citep[][]{Kormendy2001}, massive black hole pairs (MBHPs) and, at even smaller separations, massive black hole binaries (MBHBs) are predicted to form in the aftermath of the merger \citep[][]{Begelman1980}.

In our current view, the two MBHs, still unbound from one another, residing at the centres of the merging galaxies, are expected to drift toward the centre of the newly formed galaxy because of dynamical friction. The dynamical friction efficiency at large MBH separations ($0.001-10$ kpc)  is still poorly constrained \citep[see][]{Dotti2012}, due to both the possible interaction between the MBHs and galaxy substructures such as spiral arms, bars and stellar or gaseous clumps \citep[e.g.][]{Fiacconi2013, delValle2015, SouzaLima2020, Bortolas2020, Bortolas2022} and tidal effects. More specifically, tides can significantly increase the pairing timescale by decreasing the mass of the progenitor galaxy cores,\footnote{The efficiency of tidal stripping depends on the possible deepening of the potential well of the core, as a consequence of merger driven gas inflows \citep[e.g.][]{Callegari09}.} \citep[e.g. ][]{Governato94, Pfister19, Varisco24}.

Past this stage, the evolution is expected to be driven by three-body scatterings with stars \citep[][]{Begelman1980, Mikkola92} and/or the interaction with the gas surrounding the two MBHs \citep[][]{Armitage02, Haiman09}.
If the abovementioned processes manage to efficiently remove orbital energy and angular momentum from the two MBHs, the resulting binary, i.e. the system when the two MBHs are gravitationally bound to each other, will coalesce due to gravitational wave emission \citep[see][]{Thorne1976}, becoming one of the most promising source population to be detected by space-borne interferometers \citep[e.g., the Laser Interferometer Space Antenna, LISA,][]{Amaro-Seoane2023} and pulsar timing arrays \citep[PTAs, see][]{Verbiest2016}. Recently, evidence for a gravitational wave background possibly generated by inspiralling MBHBs has been found in PTA data \citep[see][]{Agazie2023, Reardon2023, Xu2023, antoniadis2024, Miles2025}.
Even though the background is likely generated by MBHBs,  the uncertainties in stellar and gas-driven evolution result in a significant scatter in the predicted GW signal \citep[e.g.][and references therein]{Varisco2021, Franchini23}. Stringent EM constraints on MBHPs and MBHBs are needed for a comparison between theory and observation.

To date, there are only $\sim100$ confirmed observations of MBHPs resolved as dual AGN, see \citealt{De_Rosa2019} for a review, \citealt{Zhang_2021} for a more recent collection of $\sim 10$ kpc MBHPs, \citealt{Hwang20, Manucci2022, Scialpi2024, Chen23, Wu24, Chen24, Schwartzman24} for recent searches of $\sim$ kpc MBHPs and \cite{Trindade24} for a candidate with $\sim100$ pc separation.

A secure identification of spatially unresolved MBH pairs and binaries is challenging. 
Current searches for unresolved signatures \citep[see][for a review]{DOrazio2023} have focused on close MBHBs where quasi-periodic modulations of the continuum \citep[][]{Valtonen08, Ackermann15, Graham2015,  DOrazio2015, Li2016, Charisi16, Sandrinelli16, Sandrinelli18, Severgnini18, Li+2019, LiuGez+2019, Hu2020, Chen+2020, Zhu2020, Cocchiararo2024, rigamonti2025b, bertassi2025identificationperiodicitiesarbitraryshapes} and of the broad emission line (BEL) profile \citep[][]{Shen2010, Tsalmantza2011, Eracleous2012, Decarli2013, Runnoe2017, Runnoe25, sottocorno2025, Rigamonti2025_PG1302-102, Bertassi2025} are expected to occur on timescales of months to centuries.

At large enough separations, the search for photometric and spectroscopic features becomes either impossible (as the underlying assumptions of the models break) or unfeasible (as the required length of the observational campaigns to observe such modulations becomes too long).
One exception to this is a test originally proposed by \cite{Gaskell88} and recently quantitatively detailed by \cite{Dotti2023},  which assumes that the two MBHs are far enough from each other for their emission (both in the continuum and in the BELs) to be uncorrelated on observational timescales. In this case, different spectral regions of BELs react at two independent continua. This test allows for the identification of pairs and binaries with orbital periods of decades to centuries using reverberation-mapping campaigns lasting $\lsim 1$ yr \citep{Dotti2023}, but fails when the separation is so large (e.g. $\sim 0.5$ pc for a binary of $\sim 10^6 \msun$) that the two independent BLRs have a too-small relative shift, due to the small orbital velocity of the binary \citep[$\lsim 100$ km/s, see][]{Dotti2023}.

In this work, we present a new photometric test for couples of active MBHs (either MBHBs or MBHPs) too widely separated to be identified through continuum periodicity (e.g., with separations $\gsim 0.001$ pc for a binary of $\sim 10^6 \msun$), but too close to be spatially resolved. For example, assuming an angular resolution of $\approx 0.65 \ \rm{arcsec}$ in $r$-band for the Vera Rubin observatory \citep{LSST}, the maximum projected separation between the two MBHs for the test to succeed is $\approx 5$ kpc at $z\approx 1$  and $\approx 1.2$ kpc at $z\approx 0.1$. In the following, we will refer to such systems as unresolved black hole duos (UBHDs).

The test leverages the intrinsic variability of AGN light curves \citep[see][for a review]{paolillo2025} well modelled with a damped random walk \citep[DRW, see][]{Kelly2009, Kozlowsky2010, MacLeod2010}, a stochastic process with a red-noise\footnote{Notably, this red noise presents a major limitations in the photometric periodicity searches for close MBHBs, leading to false positives \citep[][]{Vaughan2016}.}, i.e. a frequency-dependent noise with power increasing toward the lower frequencies and becoming white noise, i.e. frequency independent at the lowest frequencies(see section~\ref{sec:methodology}).

At the separations considered in this study, the short-timescale variability properties
of the two MBHs will be uncorrelated (as in the spectroscopic test proposed by \citealt{Gaskell88, Dotti2023}).
With simulations of UBHDs consisting of two DRW components, we test whether these systems can be distinguished from the typical AGN behaviour described by a single DRW. For this, we employ a Bayesian model comparison.
By allowing for the identification of unresolved UBHDs \citep[associated with long residence timescales, which are expected to be abundant, see e.g.][]{Haiman09}, the test has the potential to unveil large samples of UBHDs compared to the close MBHBs already searched for in time-domain surveys, both current (e.g., the Zwicky Transient Facility, ZTF, \citealt{Bellm_2018}; the Catalina Real-Time Transient Survey, CTRS \citealt{Djorgovski2011}) and upcoming (e.g., the Vera Rubin Observatory's 10-year Legacy Survey of Space and Time, LSST \citealt{LSST}; the Roman Space Telescope’s High Latitude Time Domain Survey, HLTDS \citealt{Roman}). 

The paper is organised as follows: in section \ref{sec:methodology}, we describe the properties of the DRW  model. We describe how mock light curves are generated and discuss the methods with which the parameters are constrained and the model comparison is performed. In section \ref{sec:results}, we discuss the fraction of false positives expected, we quantify how well the injected parameters are retrieved, and identify the regions in the parameter space in which the test correctly identifies a UBHD. Our main conclusions are presented in section \ref{sec:conclusions}. Appendix~\ref{app:periodogram_computation} discusses the properties of light-curve periodograms, highlighting their limitations that prompted the design of the analysis discussed in the main text. 
A less statistically solid test searching for UBHDs by fitting the power spectral density (PSD) obtained through the Lomb-Scargle periodogram \citep[see][and references therein]{VanderPlas2018} shape directly is discussed in appendix~\ref{app:PSD_fitting}.

\section{Methodology} \label{sec:methodology}

In this work, we focus on UBHDs, i.e. MBHs separated enough to have independent emission variability, so that the PSD of the resulting time series is given by the sum of the power spectra of the two processes. Furthermore, we assume that the only source of variability for each MBH is the intrinsic one produced by a DRW. We generate mock DRW light curves for each component MBH using \texttt{celerite}\footnote{The documentation about the package can be found at the following link: \url{https://celerite.readthedocs.io/en/stable/python/install/}} \citep[see][]{celerite} and create mock UBHD light curves by summing the realisations of the flux of two different MBHs (as detailed in section \ref{subsec:lightcurve_generation}). First, we consider evenly sampled time series, and then generalise the work to unevenly sampled time series, simulating the expected cadence of LSST. Next, we estimate the parameters and the evidence of a given model, either single MBH or UBHD, sampling the likelihood with a nested sampling algorithm \citep[see][]{Skilling2006} implemented in \texttt{Raynest}\footnote{The documentation about this package can be found at this link: \url{https://github.com/wdpozzo/raynest}} \citep[][]{CPnest}. Finally, we select the best model by comparing the statistical evidence for each model.

\subsection{Noise models} \label{subsec:Noise_models}
The behaviour of a quantity evolving under a DRW is described by the solution to the following stochastic differential equation  \citep[see][]{Kelly2009}: 
\begin{equation} \label{eq:DRW_stochastic}
    dX(t)= -\frac{1}{\tau} X(t) \ dt + \hat{\sigma} \ \sqrt{dt} \ \epsilon(t) + \ b \ dt \ \ \rm{with} \ \tau, \ \hat{\sigma}, \ b \ >0.
\end{equation}
Here $X(t)$ is the process that is being observed as a function of time, $\tau$ is the so-called damping timescale of the process (i.e. the time that is needed for the time series to become roughly uncorrelated), $\epsilon(t)$ is a white noise with zero mean and unit variance, and $\hat{\sigma}$ describes the lightcurve variations at small timescales and is related to the variance of the process $\sigma=\hat{\sigma} \tau /2$. Finally, $b$ is related to the mean of the process given by $ b\tau$.
Such a noise model has a PSD of the form:
 \begin{equation}
    P_{\rm{sing}}(f)= \frac{2 \hat{\sigma}^2 \tau^2}{1+(2 \pi \tau f)^2} \ ,
    \label{eq:DRW_power}
\end{equation}
while its covariance matrix can be written as:
\begin{equation}
     C^{\rm{sing}}_{i,j}=C^{\rm{DRW}}_{i,j}=\frac{1}{2}\hat{\sigma}^2 \tau \exp \left(-\frac{\Delta t_{i,j}}{\tau}\right)
    \label{eq:DRW_covariance}
\end{equation}
where $\Delta t_{i,j}=|t_i-t_j|$ and $i,j=1...N$, represent two arbitrary observations out of the total number of $N$ pointings. 

When considering an UBHD, assuming the light curve from one MBH to be independent of the other, the PSD can be written as the sum of the two PSDs:
\begin{equation}
    P_{\rm{UBHD}}(f)= \frac{2\hat{\sigma}_1^2 \tau_1^2}{1+(2 \pi \tau_1 f)^2} + \frac{2\hat{\sigma}_2^2 \tau_2^2}{1+(2 \pi \tau_2 f)^2}
    \label{eq:DRW_power_bin}
\end{equation}
where $\tau_1, \ \hat{\sigma}_1, \ \tau_2, \ \hat{\sigma}_2$ are the damping timescales and short-term variability magnitudes of the two processes, respectively. The covariance function for this model can be written as the sum of two covariance functions:
\begin{equation}
\begin{split}
    C^{\rm{UBHD}}_{i,j }&=C^{\rm{DRW, 1}}_{i,j }+C^{\rm{DRW,2}}_{i,j}= \\ & =\frac{1}{2}\hat{\sigma}_1^2 \tau_1 \exp \left(-\frac{\Delta t_{i,j}}{\tau_1}\right)+\frac{1}{2}\hat{\sigma}_2^2 \tau_2 \exp \left(-\frac{\Delta t_{i,j}}{\tau_2}\right)\ .
    \end{split}\label{eq:DRW_covariance_bin}
\end{equation}
This is consistent with the assumption of two independent, stationary, zero-mean processes being added, as the absence of cross-correlations between the two processes allows both the covariance functions and PSDs to be additive.

Several studies have been carried out to find a relation between the parameters of the stochastic equation and the physical parameters of the MBH \citep[see][for some examples]{Kelly2009, MacLeod2010, Burke2021, Arevalo2023, Arevalo2024, Su2024, Goncalves25}, but there is currently no consensus regarding these relationships.
The methodology followed in this work does not require any strong assumptions about the UBHD physical parameters. Since our objective is to explore the parameter space as broadly and agnostically as possible, we will refrain from assuming any relation between the parameters of the noise model and those of the MBHs.

\subsection{Time series generation}\label{subsec:lightcurve_generation}
Each light curve is generated by sampling the flux of the MBH using \texttt{celerite}. Once the kernel of the Gaussian process (GP) is chosen, the light curve can be sampled with an arbitrary sampling pattern and baseline.
In this work, the injected parameters are drawn from log-uniform distributions for the damping timescales and uniform distributions for the variances of the two processes, as follows:

\noindent $\bullet$ Single MBH case:
\begin{align*}
    &10 \ \mathrm{d} < \tau < 500 \ \mathrm{d}, \\ 
    &0 < \hat{\sigma}^2 < 1
\end{align*}

\noindent $\bullet$ UBHD case:
\begin{align*}
    &10 \ \mathrm{d}< \tau_1 < 500 \ \mathrm{d}, \\
    &10 \ \mathrm{d} < \tau_2 < \tau_1 , \\
    & 0 < \hat{\sigma}^2_1 < 1, \\
    & 0 < \hat{\sigma}^2_2 < 1
\end{align*}

\begin{figure*}[h!]
    \centering
    \includegraphics[width=\hsize]{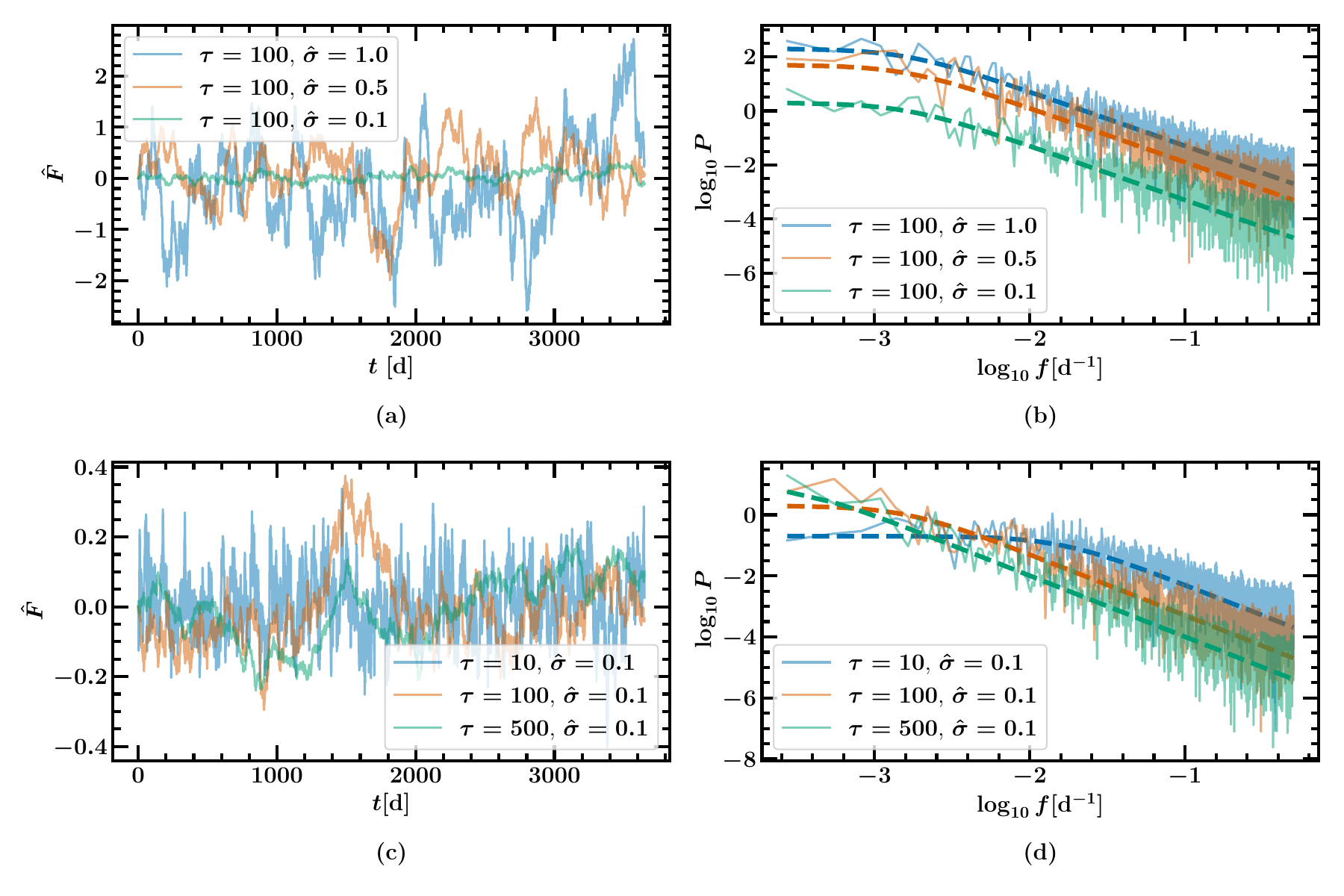}
    \caption{Examples of light curves generated using \texttt{celerite} varying the process parameters: the damping timescale $\tau$, panel (a), and the variability amplitude $\hat{\sigma}$, panel (c). In the right panels, the retrieved periodograms (computed as discussed in Appendix \ref{app:periodogram_computation}) as well as the theoretical expected PSD (dashed lines) are shown. The time series ($\hat{F}$) are centred at zero and have arbitrary units, the parameter $\sigma$ and power ($P$) should be considered in terms of the same dimensional scale as $\hat{F}$, with $P$ having units of the square of the units of $\hat{F}$.}
    \label{fig:multiplot}
\end{figure*}

In our work, we sample the time series for 10 years to resemble the LSST nominal survey. As noted in \cite{Kozlowsky2017}, to obtain good constraints on the flat part of the PSD, the process should be observed for at least $10 \ \tau$. With the assumed observational baseline, the flat part of the PSD can be observed particularly well for MBHs with a mass up to $\sim 10^8 \msun$ since they are expected to have a damping time of the order of 200 days \citep{Burke2021}. We initially sample the light curves evenly in time every 3 days, and then generalise the test to more realistic light curves by omitting three to four months of observations each year to incorporate the inevitable gaps of ground-based observations. We assume a relative flux uncertainty of $0.01$. Such an uncertainty corresponds to a magnitude of $\sim 21$ in the r-band of a single LSST visit, based on the expected photometric precision \cite{LSST}. In the case of an evenly sampled light curve, we assume a time step of $\Delta t= 3 \ \rm{day}$, roughly corresponding to the cadence of the Wide Fast Deep survey mode of LSST, which will cover about 95\%  of the survey time. Examples of light curves generated in such a way, along with their respective PSDs, are shown in Figure \ref{fig:multiplot}. Here, it is possible to see how a change in the parameters changes the observed light curve and its power spectrum. More specifically, a change in the parameter $\hat{\sigma}$ results in a uniform vertical shift of the PSD, with higher values of $\hat{\sigma}$ leading to increased power at all frequencies, as shown in (b) panel of Figure \ref{fig:multiplot}. A change in the damping timescale $\tau$ causes a change in the break frequencies with a higher $\tau$ moving the flat part of the spectrum toward lower frequencies, as shown in panel (d) of Figure \ref{fig:multiplot}. 
The light curves for the UBHD case are constructed by summing two light curves generated as in the single MBH scenario with equal weight.

\subsection{Parameter estimation and model selection} \label{sec:parameter_estimation}
In this work, we combine nested sampling and Gaussian processes to estimate the posterior distributions of the parameters and to compare the two competing models \citep[see][for works using a similar approach]{Covino_2020, Zhu2020, Covino2022, Witt_2022}. While Gaussian processes often employ a maximum likelihood approach to optimise kernel hyperparameters for computational convenience, we take a fully Bayesian approach by using nested sampling to marginalise over these parameters. More specifically, nested sampling transforms the multidimensional integral for the evidence into a one-dimensional integral over the prior volume and explores nested contours of increasing likelihood by iteratively replacing low-likelihood samples with higher-likelihood ones. Unlike Markov Chain Monte Carlo (MCMC) algorithms, nested sampling allows us to capture the full posterior distribution and compute the model evidence, enabling a statistically robust comparison between models \citep{Skilling2006, Ashton_2022, Buchner_2023}. We note that, due to the parallelism between Gaussian process kernels and PSD, UBHDs can be searched for through the fitting of the periodograms of evenly spaced light curves. However, for uneven data samplings, a consistent Bayesian analysis is not possible (as detailed in Appendix~\ref{app:periodogram_computation}). An analysis directly performed on the PSDs is presented in Appendix~\ref{app:PSD_fitting} together with its shortcomings.

The fitting of the light curve through Gaussian processes is performed using the Gaussian process regressor \texttt{celerite}, using a kernel in the form:

\begin{equation}
k_{\mathrm{single} \  i, j} = a e^{-c \ \Delta t_{i,j}} .
\label{eq:celerite_RealTerm}
\end{equation}
 Equation (\ref{eq:celerite_RealTerm}) becomes equal to the kernel in Equation (\ref{eq:DRW_covariance}) by setting $a=0.5 \tau \sigma^2 $ and $c= 1/\tau$. In the context of \texttt{celerite}, a kernel such as the one in Equation (\ref{eq:celerite_RealTerm}) is known as RealTerm. The UBHD model can then be written as the sum of two RealTerm kernels:
\begin{equation}
k_{\mathrm{UBHD}}(\Delta t) = a_{1} e^{-c_{1} \Delta t} +  a_{2} e^{-c_{2} \Delta t} .
\label{eq:celerite_RealTerm_pair}
\end{equation}
We note that we chose to use \texttt{celerite} due to its computational efficiency in evaluating the likelihood for kernels such as those presented in Equation~(\ref{eq:celerite_RealTerm}). Specifically, the computational cost of \texttt{celerite} scales linearly with the number of data points, in contrast to the cubic scaling of standard Gaussian Process methods.

The nested sampling is performed using \texttt{RayNest}. The priors are a log-uniform distribution in $\tau$ in between 10 and 500 d, and log-uniform in  $\hat{\sigma}^2$ in between $10^{-6}$ and $10^{6}$. We emphasise that the prior distribution on the variance $\hat{\sigma}^2$ is much wider than the distribution used to generate the light curves. This choice was made to ensure that the results and posteriors are not biased due to an artificial truncation by the prior boundaries, which can affect the resulting posterior and evidence.

For model comparison, we computed the Bayes Factor, defined as the ratio between the evidence of the two models:
\begin{equation} \label{eq:Bayes_factor}
    B_{\rm{pair,sing}}=\frac{\int p(D|M_{\rm{pair}},\boldsymbol{\theta}_{\rm{pair}} \ )\ p (\boldsymbol{\theta}_{\rm{pair}}|M_{\rm{pair}}) \ d\boldsymbol{\theta}_{\rm{pair}}}
    {\int p(D|M_{\rm{sing}},\boldsymbol{\theta}_{\rm{sing}} \ )\ p (\boldsymbol{\theta}_{\rm{sing}}|M_{\rm{sing}}) \ d\boldsymbol{\theta}_{\rm{sing}}}
\end{equation}
where $p(D|M,\boldsymbol{\theta},I )$ is the likelihood function and $ p (\boldsymbol{\theta}|M,I)$ is the prior, $D$ is the data, i.e. the observed light curve ($t_i$, $F_i$), with i,1,2,...N the number of data points and $F_i$ being the flux observed at time $t_i$, $M$ is the considered model, and finally $\boldsymbol{\theta}$ are the parameters for this particular model. In this work, we assume that the two models have the same prior probability

In the particular case of Gaussian processes, the likelihood of the data is given by \citep[see][]{Rasmussen_2006}:
\begin{equation}
p(D \mid M, \boldsymbol{\theta}, I) = 
\frac{1}{(2\pi)^{\frac{N}{2}} |\boldsymbol{\Sigma}|^{\frac{1}{2}}} \exp\left(
-\frac{1}{2} (D - \boldsymbol{\mu})^{T} \boldsymbol{\Sigma}^{-1} (D - \boldsymbol{\mu})
\right)
\label{eq:GP_likelihood}
\end{equation}
where $\boldsymbol{\mu}$ is the mean function of the Gaussian Process, $N$ is the number of observations, and $\boldsymbol{\Sigma}$ and $|\boldsymbol{\Sigma}|$ are the covariance matrix of the GP and its determinant.

Here, we consider the evidence for UBHD to be significant when $B_{\rm{pair,sing}}>3$, i.e. if it shows at least substantial evidence according to the Jeffreys scale \citep[][]{jeffreys1998}.

\section{Results} \label{sec:results}
The test proposed in this work aims to assess the presence of a UBHD directly from the analysis of the observed light curve. To achieve this, we analyse realistic time series using Gaussian processes and nested sampling.

In this section, we present the main findings of our study. First, we assess the false positive rate of the proposed signature by analysing single MBH light curves with both models and computing the corresponding Bayes factor (Section~\ref{sec: false_positives}) to infer the percentage of single MBH light curves wrongly identified as UBHDs.

Next, we evaluate how accurately the nested sampling procedure recovers the injected parameters (Section~\ref{sec:parameter_recovery}), considering both the single MBH and UBHD scenarios, and using both evenly and unevenly sampled light curves.

Finally, we explore which regions of the model parameter space allow for a reliable identification of UBHDs. To do this, we randomly sample parameters and evaluate the Bayes factor to determine the conditions under which the test is most effective (Section~\ref{sec: best_regions}).

\subsection{False positives} \label{sec: false_positives}
To quantify the robustness of the proposed signature, it is important to quantify how unique it is versus how often it can be confused with the typical DRW variability of a single MBH. For this, we assess the fraction of false positives produced by the test by simulating 1000  evenly sampled and 1000 unevenly sampled single MBH light curves and calculate the Bayes Factor for the UBHD  vs single MBH model, as described in section \ref{sec:parameter_estimation}. We then compute the fraction of light curves that our test identifies as UBHDs, i.e. the fraction of light curves for which the Bayes factor is bigger than 3. We find the $0.2\%$ and $0.59\%$ of false positives in the evenly and unevenly sampled scenarios, respectively. The percentage of false positives remains smaller than $1\%$, increasing when the sampling becomes uneven, even if the length of the observational campaign remains the same. We emphasise that the overall false detection rate is very low (regardless of light curve quality).

\subsection{Parameter recovery} \label{sec:parameter_recovery}
Before searching for the best parameter space in which the analysis of the photometric light curve can show evidence for the presence of an UBHD, we test how well our algorithm can recover the parameters of the injected light curve for each model. This comparison is not a simple consistency check between model and data realisations, as the processes we are looking at are intrinsically stochastic, i.e. individual realisations would differ even for the same set of parameters. To assess the goodness of the parameter estimation, we compute the relative error between the injected $\theta_{\mathrm{inj}}$ and recovered $\theta_{\mathrm{rec}}$ parameters as: 
\begin{equation}
    \delta \theta =\frac{\theta_{\mathrm{rec}}- \theta_{\mathrm{inj}}}{{\theta_{\mathrm{inj}}}}\times 100 \%
    \label{eq:rel_err}
\end{equation}
and the relative interquantile range $\sigma_{\theta}$ defined as:
\begin{equation}
\sigma_{\theta}=\frac{q_{\theta,0.84}-q_{\theta,0.16}}{\theta_{\rm{inj}}} \times100\%
\label{eq:iqr_range}
\end{equation}
where $\theta_\mathrm{inj}$ is one of the parameters with which the light curve has been generated, $\theta_\mathrm{rec}$ is the median value of the posterior distribution for that particular parameter, $q_{\theta,0.16}$ and $q_{\theta,0.84}$ are respectively the 0.16 and 0.84 quantiles of the posterior distribution. 

\begin{figure*}[ht!]
    \centering
    \includegraphics[width= \hsize]{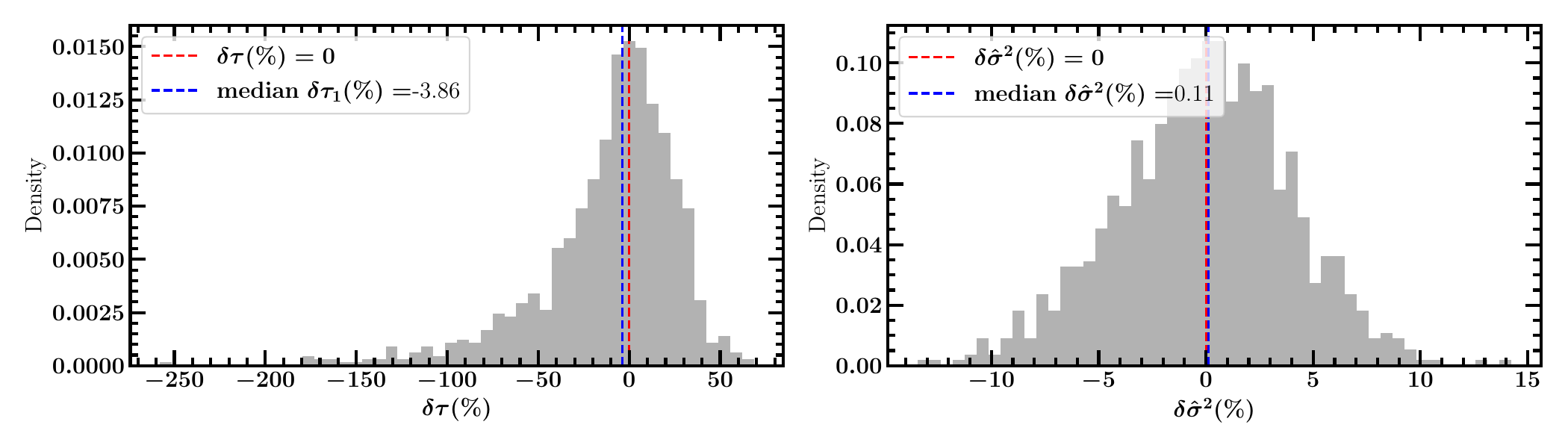}
    \caption{Distribution of relative errors for the retrieved parameters in the single MBH evenly sampled} case. 
    \label{fig:delta_sigma_dist}
\end{figure*}

In Figure~\ref{fig:delta_sigma_dist}, we show the distribution of the relative errors $\delta\theta$ for 1000 evenly sampled light curves for the single MBH case. The dashed vertical red line indicates the optimal scenario $\delta\theta=0$ while the dashed vertical blue line indicates the median value of $\delta\theta$. We see that both parameters are recovered well as these distributions are scattered around $\delta\theta=0$ for both $\tau$ and $\hat{\sigma}$. The values we found for the median relative error $\delta\theta$ and relative interquantile range $\sigma_{\theta}$ for both the even and uneven scenarios are reported in Table~\ref{tab:par_estimate_sing}.
\begin{table}[h!]
    \centering
    \caption{Parameter recovery for the single MBH model.}
    
    \begin{tabular}{|c|c|c|}
    \hline
        & Even sampling (\%)& Uneven sampling (\%)  \\
        \hline
      $\delta\tau$  & -3.86 & -4.13\\
      \hline
      $\delta\hat{\sigma}$  & 0.11  & 0.03 \\
      \hline
      $\sigma_{\tau}$ & 82.76 & 84.18   \\
      \hline
      $\sigma_{\hat{\sigma}}$  & 0.04 & 0.04\\
      \hline
    \end{tabular}
    \tablefoot{Median values of the relative error $\delta\theta$ and relative interquantile range $\sigma_{\theta}$ for the parameters $\tau$ and $\hat{\sigma}$ of the single MBH model.}
    \label{tab:par_estimate_sing}
\end{table}

\begin{figure*}[ht!]
    \centering
    \includegraphics[width= \hsize]{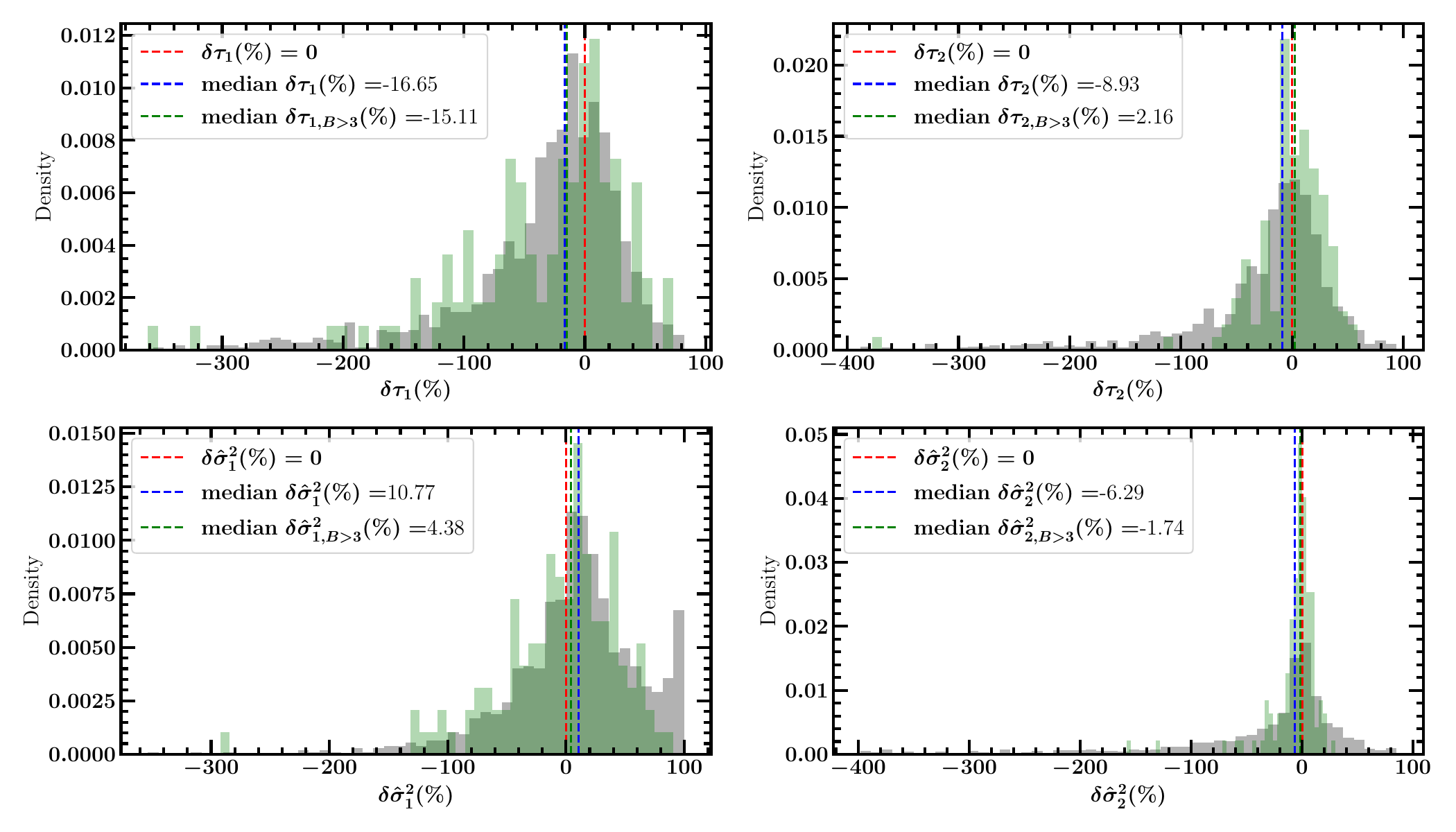}
    \caption{Distribution of relative errors for the retrieved parameters in the  UBHD evenly sampled case. In green, the distribution for the cases in which the UBHD model is favoured.}
    \label{fig:deltas_bin}
\end{figure*}

For the damping timescale, we see a tail at lower values, due to the fact that the length of the observational campaign is not long enough to efficiently sample the flat part of the PSD. In Figure \ref{fig:deltas_bin}, we do the same for the UBHD case. Here, the scatter from the true value is higher than that in the single DRW scenario, as can be seen from the values of the median relative error $\delta\theta$ and relative interquantile range $\sigma_{\theta}$ reported in Table~\ref{tab:par_estimate_UBHD}. 

\begin{table}[h!]
    \caption{Parameter recovery for the UBHD model.}
    \centering
    \begin{tabular}{|c|c|c|}
    \hline
        & Even sampling (\%) & Uneven sampling (\%)  \\
        \hline
      $\delta\tau_1$  & -16.65 & -13.31\\
      \hline
      $\delta\hat{\sigma_1}$  & 10.77 & 12.15 \\
      \hline
      $\delta\tau_2$  & -8.93 & 4.20 \\
      \hline
      $\delta\hat{\sigma_2}$  & -6.29  & -0.73 \\
      \hline
      $\sigma_{\tau_1}$  & 260.56 & 266.95 \\
      \hline
      $\sigma_{\hat{\sigma_1}}$  & 9.93 & 52.54 \\
      \hline
      $\sigma_{\tau_2}$  & 103.23 & 77.98\\
      \hline
      $\sigma_{\hat{\sigma_2}}$  & 6.08 & 18.01 \\
      \hline
    \end{tabular}
    \tablefoot{Median values of the relative error $\delta\theta$ and relative interquantile range $\sigma_{\theta}$ for the parameters $\tau_1$, $\tau_2$, $\hat{\sigma}_1$$\hat{\sigma}_2$ of the UBHD model.}
    \label{tab:par_estimate_UBHD}
\end{table}

The green distributions in Figure~\ref{fig:deltas_bin} represent the distributions of $\delta\theta$ for the combination of parameters where the test correctly identifies a UBHD. The median values of the relative errors and the relative interquantiles range of such light curves are reported in Table~\ref{tab:par_estimate_UBHD_good}.
\begin{table}[h!]
    \caption{Parameter recovery for the correctly identified UBHD light curves}
    \centering
    \begin{tabular}{|c|c|c|}
    \hline
        & Even sampling (\%) & Uneven sampling (\%)  \\
        \hline
      $\delta\tau_1$  & -15.11 & -19.09\\
      \hline
      $\delta\hat{\sigma_1}$  & 4.38 & 10.44 \\
      \hline
      $\delta\tau_2$  & 2.16 & 2.89\\
      \hline
      $\delta\hat{\sigma_2}$  & -1.74  & -2.42 \\
      \hline
      $\sigma_{\tau_1}$  & 155.52 & 160.09 \\
      \hline
      $\sigma_{\hat{\sigma_1}}$  & 3.84 & 4.08 \\
      \hline
      $\sigma_{\tau_2}$  & 45.10 & 50.75\\
      \hline
      $\sigma_{\hat{\sigma_2}}$  & 0.69 & 0.63 \\
      \hline
    \end{tabular}
    \tablefoot{Median values of the relative error $\delta\theta$ and relative interquantile range $\sigma_{\theta}$ for the parameters $\tau_1$, $\tau_2$, $\hat{\sigma}_1$$\hat{\sigma}_2$ of the UBHD light curves correctly identified by the test (the green distributions in Figure~\ref{fig:delta_sigma_dist}).}
    \label{tab:par_estimate_UBHD_good}
\end{table}
In this case, one can see that the scatter overall becomes smaller, but with $\tau_1$ still showing wide tails in the distributions, similarly to the single DRW case.

In Table~\ref{tab:percentages_total}, we report the percentage of light curves for which all the parameters are retrieved with a relative error within the $10\%$, $20\%$ and $50\%$ for the single MBH, UBHD and correctly identifies (green distributions in Figure~\ref{fig:deltas_bin}) cases, respectively.
\begin{table*}[h!]
    \centering
    \caption{Percentages of light curves for which all the parameters are retrieved with a relative error within the $10\%$, $20\%$ and $50\%$}
    \begin{tabular}{|c|c|c|c|c|c|c|}
    \hline
    & single MBH (E) [\%] & single MBH (U) [\%] & UBHD (E) [\%]& UBHD (U) [\%]& UBHD \ding{72} (E) [\%] &  UBHD \ding{72} (U) [\%] \\
    \hline
    $10\%$ &  29.55 & 27.69 & 1.10 & 0.67 & 2.38 & 1.43\\
    \hline
    $20\%$ & 51.16  & 50.90 & 7.13 & 5.35 & 14.29 & 8.31 \\
    \hline
    $50\%$ & 84.92 & 86.15 &  35.14 & 20.74 & 48.41 & 39.54 \\
    \hline
    \end{tabular}
    \tablefoot{From left to right: percentages of light curves with both retrieved parameters relative errors within the $10\%$, $20\%$ and $50\%$ in the single MBH scenario, UBHD scenario and for UBHD light curves that are correctly identified (green distribution in Figure \ref{fig:deltas_bin}, indicated in the Table with UBHD \ding{72}), respectively. E and U in the parentheses indicate the even and uneven sampling, respectively.}
    \label{tab:percentages_total}
\end{table*}

\subsection{True positives} \label{sec: best_regions}
Finally, we searched for the best region of the parameter space for which the test can efficiently identify UBHDs. More specifically, we sample 1500 light curves (for both evenly and unevenly sampled light curves)varying the ratio between the two damping timescales $q_{\tau}=\tau_2/\tau_1$ with $\tau_2 <\tau_1$ and the ratio between the amplitudes of the two processes $q_{\sigma}=\sigma_2/\sigma_1=(\hat{\sigma}_2\sqrt{\tau_2})/(\hat{\sigma}_1\sqrt{\tau_1})$ with $\sigma_2 < \sigma_1$. For each pair of $q_{\tau}$ and $q_{\sigma}$, we compute the Bayes factor in order to highlight where the test works the best, i.e. where the Bayes factor is bigger than 3.

\begin{figure}
    \centering
    \includegraphics[width=\hsize]{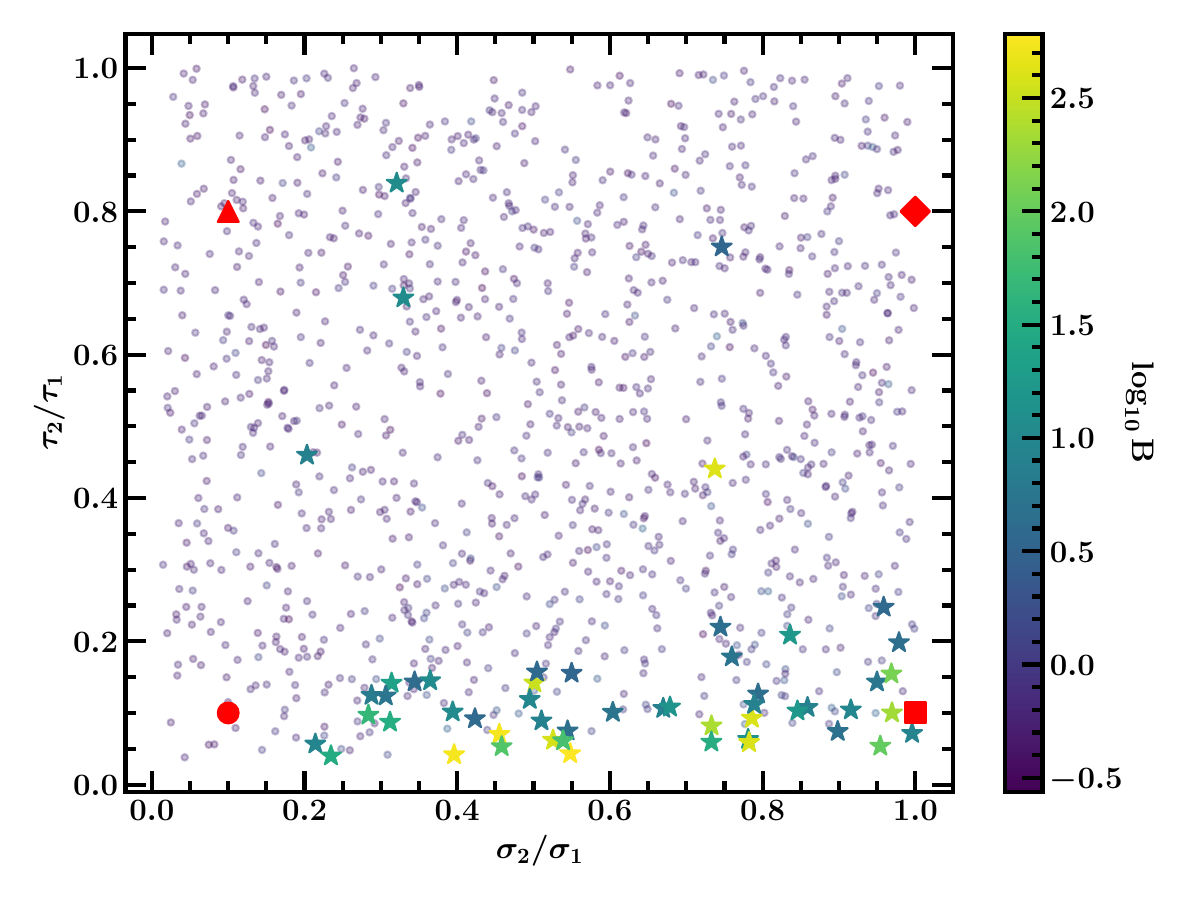}
    \caption{Map of Bayes factors for 1500 unevenly sampled realisations of UBHD light curves with a 10-year baseline. The colour scale represents the base-10 logarithm of the Bayes factor, while the axes show the ratio of the model parameters between the two light curves. The red square, diamond, triangle and circle correspond to the light curves used to compute the periodograms shown in panels a,b,c, and d of Figures~\ref{fig:double_kernel_PSD} and~\ref{fig:PSD_four_panel}, respectively.}
    \label{fig:even_3000}
\end{figure}

In Figure \ref{fig:even_3000}, we show the result in the case of unevenly sampled light curves observed for 10 years with a cadence of 3 d.   The points are colour-coded by the value of $\log_{10} B$, and the set of parameters for which the test yields a Bayes factor greater than 3 is represented by stars. 
The result does not significantly change when considering evenly sampled light curves. More specifically, in the evenly sampled scenario, $4.65\%$ of all the considered light curves are correctly identified as UBHDs, while in the unevenly sampled scenario, this fraction reduces to $3.77\%$.

From Figure \ref{fig:even_3000}, it is clear that the test we propose works only in a small region of the parameter space where $q_{\tau}\lsim 0.2$ and $q_{\sigma}>0.2$. In principle, setting a lower Bayes factor threshold could increase the fraction of correctly retrieved UBHD light curves and expand the region of the parameter space where the test works. For example, thresholds on B of 2, 1, and 0.5 would increase the fraction of correctly retrieved UBHD light curves from $3.77\%$ (when the threshold is set to $B=3$) to $5.3\%$,$ 11.3\%$ and $73.9\%$ in the case of unevenly sampled light curves and from  $4.6\%$ (for $B=3$) to $5.4\%$, $12.5\%$ and $87.6\%$ in the case of evenly sampled light curves. The ratio between the damping timescales below which the test works moves from $q_{\tau} = 0.21$ for $B=3$ to $q_{\tau} = 0.31$, $q_{\tau} \sim 0.65$ and $q_{\tau} \sim 0.87$ for B=2, 1, and 0.5, respectively. At the same time, lowering the threshold would increase the false positive fraction. For the uneven sampling scenario, the false positive fraction would increase from $0.59\%$ (for B=3) to $1\%$, $6 \%$ and $82\%$, while in the even sampling scenario the false positives would increase from $0.59\%$ (for $B=3$) to $1.07\%$, $6\%$, and $82\%$ for Bayes factor thresholds of 2, 1, and 0.5, respectively. Therefore, lower thresholds would make the statistical evidence in favour of the UBHD scenario less robust. For this reason, we choose a more conservative limit for the threshold.

\begin{figure}
    \centering
    \includegraphics[width=\hsize]{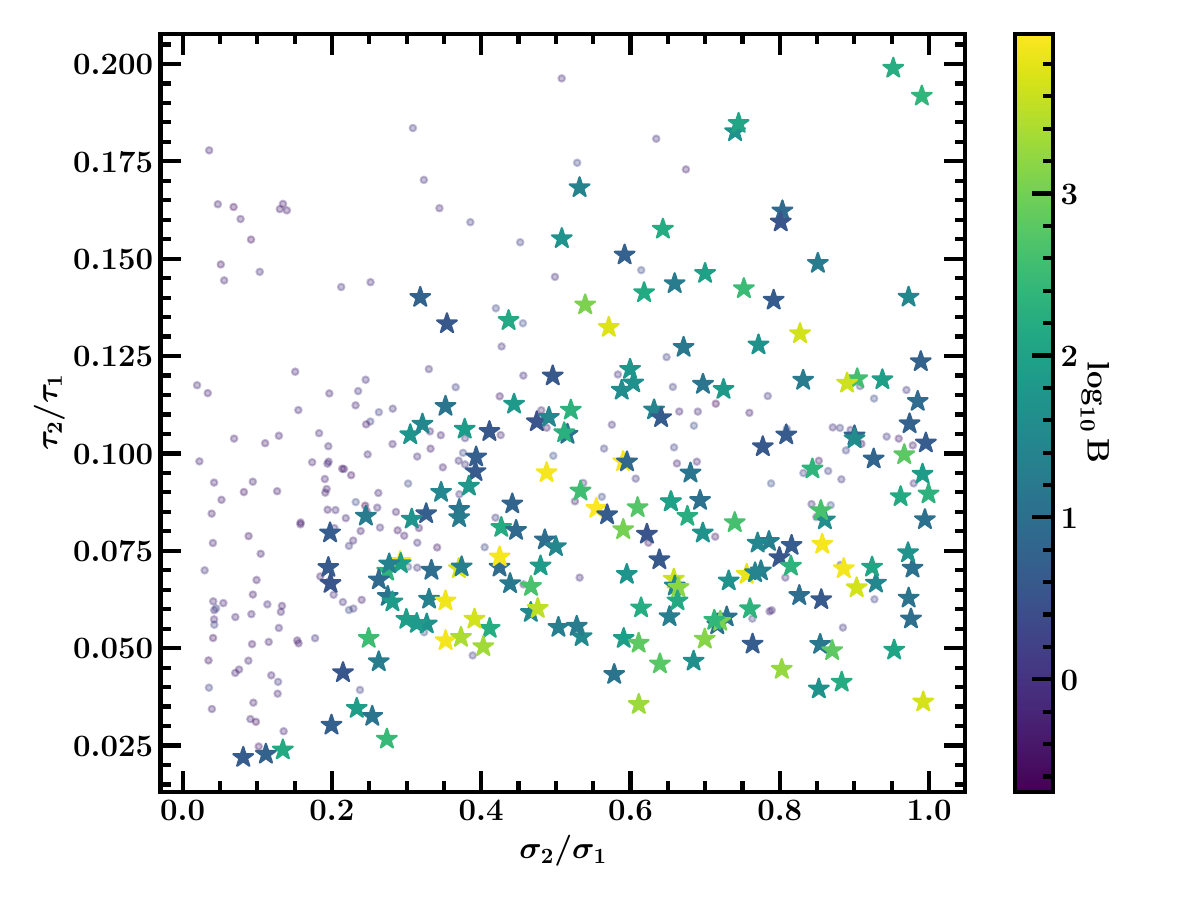}
    \caption{Same kind of map as reported in Figure \ref{fig:even_3000} for 580 light curves in the region of the parameter space where the UBHDs can be correctly identified.}
    \label{fig:even_3000_zoom_in}
\end{figure}

\begin{figure*}[h!]
    \centering
    \includegraphics[width=\textwidth]{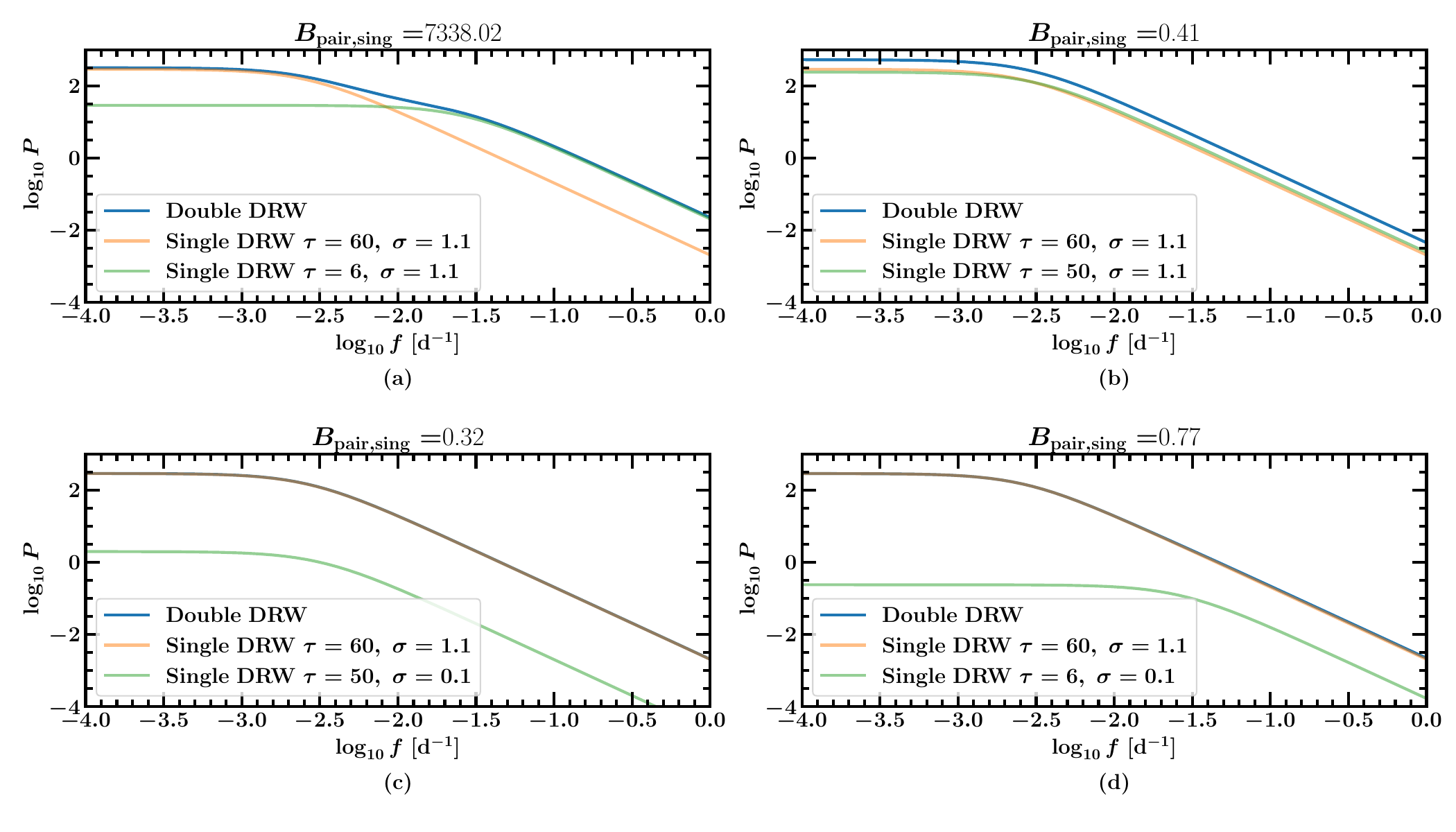}
    \caption{PSD of double DRW kernels for different parameter ratios. The Bayes ratio resulting from the GP fitting is reported in the title. In all the panels, the light curves are sampled every 3 d for 10 yr. The power is reported in arbitrary units.  
    Cases: \textbf{(a)} similar variability amplitudes but considerably different damping timescales, the test can correctly identify the UBHD (shown as the red square in Figure~\ref{fig:even_3000});  
    \textbf{(b)} similar amplitudes and similar damping timescales (shown as the red diamond in Figure~\ref{fig:even_3000});  
    \textbf{(c)} considerably different amplitudes but similar damping timescales (shown as the red triangle in Figure~\ref{fig:even_3000});  
    \textbf{(d)} considerably different amplitudes and damping timescales (shown as the red circle in Figure~\ref{fig:even_3000}).}
    \label{fig:double_kernel_PSD}
\end{figure*}

In Figure \ref{fig:even_3000_zoom_in}, we show 500 new simulations to provide a zoom-in on this region of parameter space for UBHD light curves. 
In particular, UBHDs are identified when the two MBHs have very different damping timescales and similar variability amplitudes. This fact is not surprising if one considers the shape of the PSD. In Figure~\ref{fig:double_kernel_PSD}, we show the PSDs for four cases, one of which is identified (shown in panel a) as a UBHD. More specifically, the search identifies a UBHD if, after the first plateau of the PSD up to the break frequency related to the longer damping timescale, a second plateau starts appearing up to the second break frequency related to the shorter damping timescale. 

Since the damping timescale positively correlates with the MBH luminosity as well as with the mass of the MBH, we can conclude that pairs where the two MBHs have very different luminosities or very different masses \citep[see][]{MacLeod2010}, i.e. a small luminosity ratio or a small mass ratio, are expected to be more easily identified.
The ability of this analysis to identify UBHDs large-scale photometric surveys (which provide unevenly sampled data) makes it relevant for currently available photometric surveys, most of which feature gaps and uneven pointings. 

\begin{figure}[h!]
    \centering
    \begin{subfigure}
        \centering
        \includegraphics[width=\linewidth]{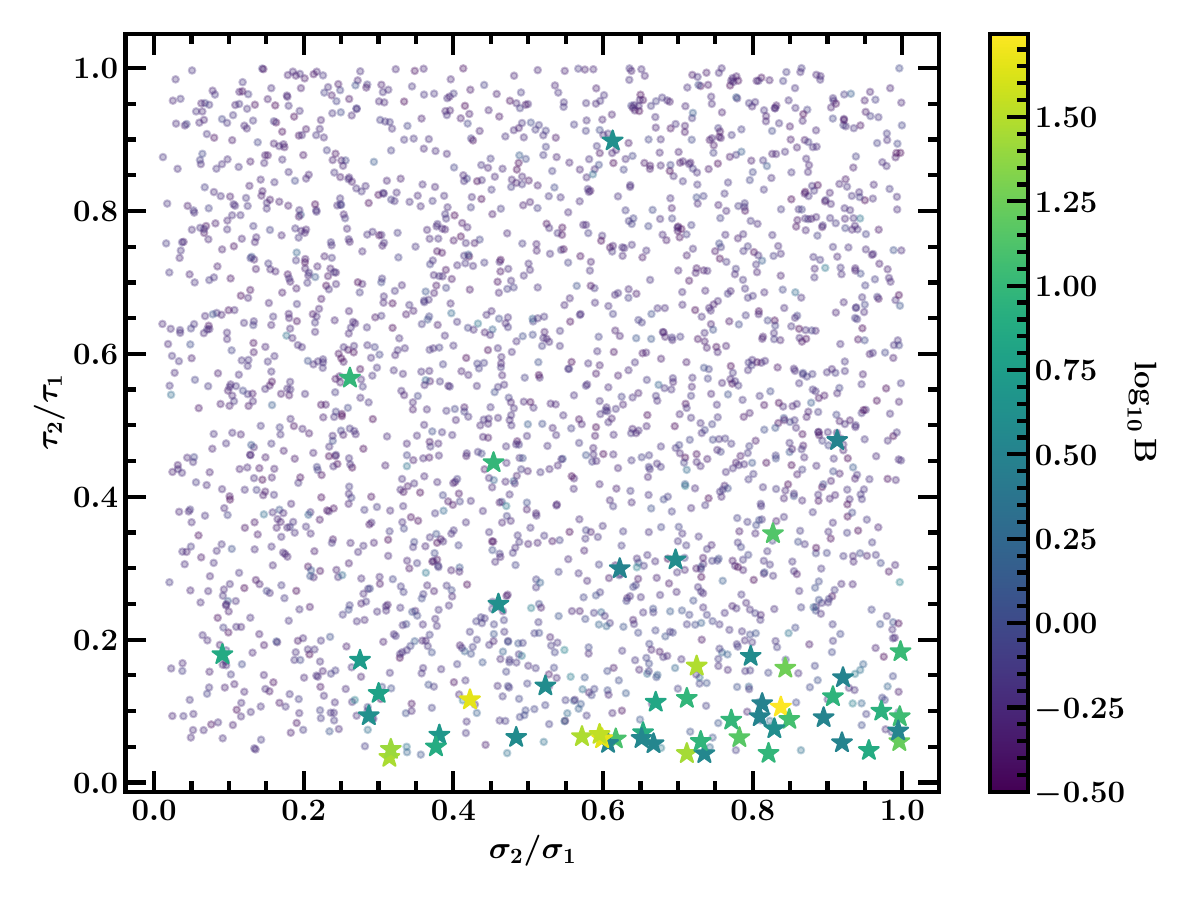}
    \end{subfigure}

    \begin{subfigure}
        \centering
        \includegraphics[width=\linewidth]{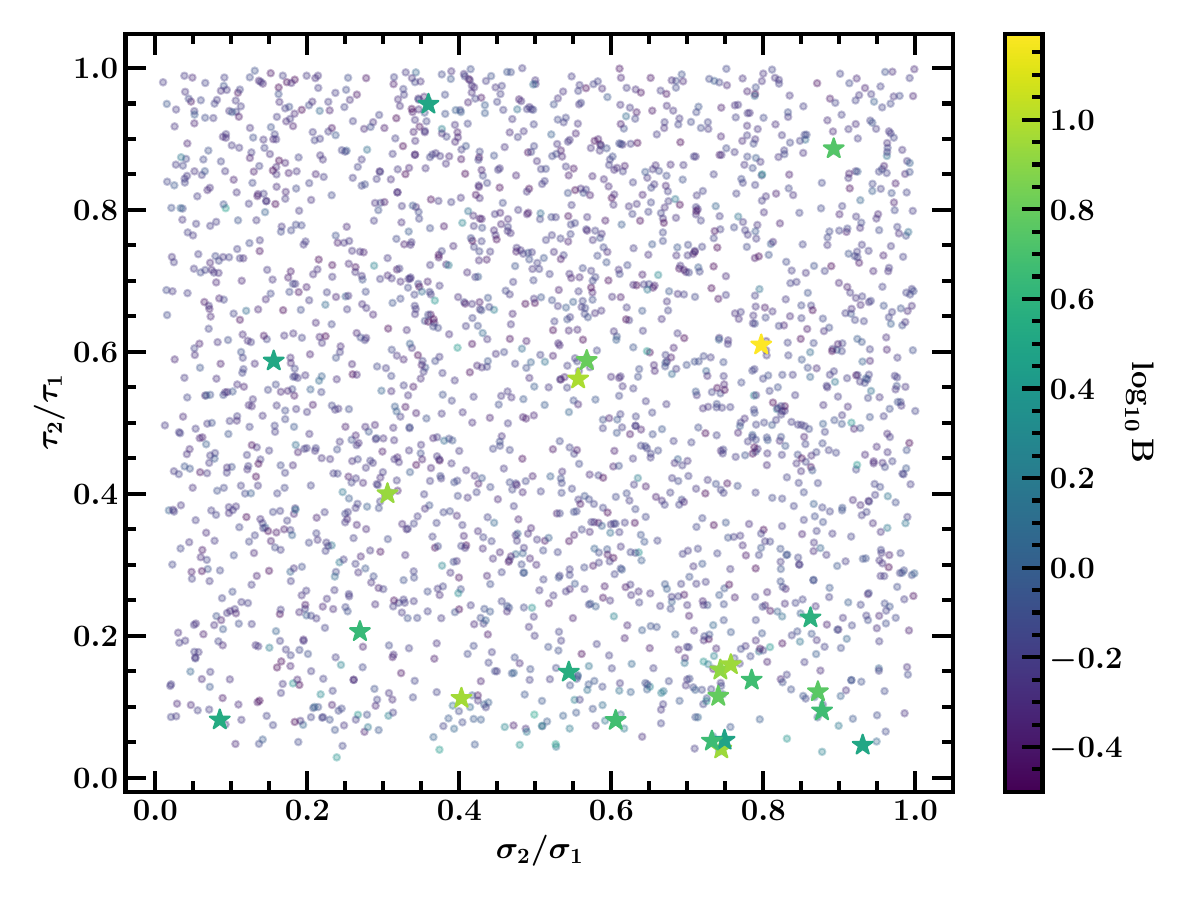}
    \end{subfigure}
    
    \caption{Maps similar to the one shown in Figures \ref{fig:even_3000} with different lengths observational baselines. In the upper panel, the light curves are observed for 6 years, while in the lower panel, the light curves are observed for 3 years. }
    \label{fig:random_maps_6_3}
\end{figure}

We repeated the same procedure, changing the length of the observational baseline. More specifically, we studied the cases of light curves observed for 6 and 3 years, keeping the uneven sampling discussed in Section \ref{subsec:lightcurve_generation}. These baselines are relevant for current time-domain surveys, e.g. ZTF, and for the early years of LSST. As can be seen by  Figure \ref{fig:random_maps_6_3}, as the length of the observation decreases, the region where the test is able to correctly retrieve the presence of an UBHD shrinks to a region of higher $q_{\sigma}$. Specifically, by repeating simulations in the region with $q_{\tau}<0.2$, analogous to those used to generate Figure~\ref{fig:even_3000_zoom_in}, we find that the value of $q_{\sigma}$ above which the $90\%$ of the correctly retrieved light curves lie moves from $q_{\sigma}=0.21$ for the $10 \ \rm{yr}$ baseline to $q_{\sigma}=0.36$ and $q_{\sigma}=0.40$ for baselines of 6 and 3 yr respectively.

\section{Discussion and Conclusions} \label{sec:conclusions}
In this paper, we studied the possibility of detecting the presence of a UBHD using AGN light curves. In particular, we examined whether we can distinguish the presence of such a system from the typical variability of an AGN with a single MBH (typically described with a DRW model).  
For this signature, we focused on the optical variability, where we will soon have large samples of quasar light curves.

We generated both single MBH mock light curves and UBHD mock light curves by summing two single MBH time series with different cadences (either evenly sampled light curves with a 3-day cadence or including observational gaps of a few months every year) and different lengths in terms of observation time (10, 6 or 3 yr). We then modelled the simulated time series employing Gaussian processes, using either a single DRW kernel or the sum of two uncorrelated DRW kernels. Finally, we checked whether the presence of a double DRW could be effectively observed through Bayesian model comparison. In particular, we used nested sampling to compute the evidence of the two models (single and double DRW) and calculated the Bayes factor, which allows us to assess which model describes the simulated data better.

In section \ref{sec: false_positives}, we studied how many false positives, i.e. light curves generated as single MBH, are better described by the double DRW model according to the Bayes factor. To do so, we generated single MBH light curves with different cadences, and for each of the time series, we computed the Bayes ratio. We considered as a false positive a time series which presented a Bayes ratio bigger than 3, i.e. the Bayes ratio indicates substantial evidence in favour of the double DRW model following the Jeffreys scale. More specifically, considering $1000$ evenly sampled light curves with a 10 yr baseline show the $0.2\%$ of false positives, while the same number of unevenly sampled light curves with the same baseline length show the $0.59\%$, demonstrating that the proposed signature has a low confusion rate with the typical AGN variability.

In section \ref{sec:parameter_recovery}, we assessed whether nested sampling successfully recovered the injected parameters, both for the single MBH scenario and for UBHDs. To do so, we computed the relative per cent error $\delta \theta$  (i.e. the normalised difference between the injected and recovered parameter, see Equation~\ref{eq:rel_err}) and the relative per cent interquantile range (i.e. the normalised interquantile range between by the injected parameter, see Equation~\ref{eq:iqr_range}).
We showed that $51.16\%$ of evenly sampled single MBH light curves, $50.90 \%$ of unevenly sampled single MBH light curves, $7.13\%$ of evenly sampled UBHD light curves and $5.35 \%$ of unevenly sampled UBHD light curves with a 10 yr baseline show recovered parameters within $20\%$ of the injected parameters.

In section \ref{sec: best_regions}, we studied which region of the parameter space allows for the detection of an UBHD. In particular, we generated light curves with different ratios between the two damping timescales and variances of the two DRW and constructed maps that show which combinations of these two ratios allow for substantial evidence for the double DRW model. We found that our test is able to identify UBHDs that show very different damping timescales, $q_{\tau}<0.2$, and similar variability amplitudes $q_{\sigma}>0.2$ for both evenly and unevenly light curves observed for 10 years. For the correctly identified UBHDs, the fraction of light curves with $\delta\theta<20\%$ grows to $14.29\%$ ($8.31\%$) for the evenly (unevenly) sampled data.
The small ratio in the damping timescales implies that only UBHDs with very different luminosities (and masses) will be correctly classified. Furthermore, we changed the length of the observation, maintaining an uneven sampling, finding that shortening the length of the observational baseline increases the $q_{\sigma}$ needed to correctly identify a UBHD.

In the appendices, we explored another possible method to detect the presence of an unresolved UBHD by studying the shape of the periodogram applicable only for evenly sampled data, which is not readily available. In Appendix \ref{app:periodogram_computation}, we showed how the periodogram can be computed and discussed the statistical properties of the classical periodogram. In Appendix \ref{app:PSD_fitting}, we showed that by using the Whittle likelihood \citep[see][]{Whittle51} to describe the distribution of the retrieved periodogram with respect to the true PSD, one can assess whether a single or double DRW better describes the observed periodogram shape. This method works only in the case of evenly sampled time series. In the case of unevenly sampled time series, this approach cannot be used as the statistical properties are different from the classical ones and depend on the particular noise considered.

Once candidates are identified, confirming the UBHD nature of objects with intermediate values of $B_{\rm{pair, sing}}$ and discerning whether the two MBHs are bound in a binary or not will require follow-up observations. High resolution imaging would have the chance to directly identify MBHPs \citep[e.g.][]{Hwang20, Manucci2022, Scialpi2024, Chen23}, while optical spectroscopy allows for checking for the presence of two sets of narrow emission lines (NELs), supporting the UBHD scenario \citep[e.g.][]{Comerford2009}. If a single set of NEL is observed, the identification of BELs that are asymmetric or offset with respect to the host rest frame in optical-UV spectra would be an indication of the presence of a close binary \citep[e.g.][]{Shen2010, Tsalmantza2011, Eracleous2012}. In such a case, reverberation mapping studies could be used to further support binary interpretation, searching for uncorrelated variability of different parts of single BELs \citep{Gaskell88, Dotti2023}.

It should be stressed that our analysis could identify systems composed of two MBHs whose hosts are not in interaction, as, for example, chance superpositions of two AGN at different redshifts or within a galaxy cluster \citep[e.g.][]{DottiRuszkowski10}. This particular scenario can be tested by looking at the frequency shift between two sets of NELs and by looking at the local density of galaxies in the UBHD--candidate neighbourhood \citep[e.g.][]{decarli14}. Another scenario that could be similar to the one explored in this work is the case of a strongly lensed quasar with angularly unresolved multiple images. However, such a scenario cannot produce false UBHDs as the light curves of the different images generated by gravitational lensing will have the same damping timescale, generated by the same quasar. 

The only differences between the lensed light curves would be the amplitude, due to the different magnification of the images, and a time delay. To test this scenario, we simulated 2000 unresolved lensed quasar light curves by simulating realisations of a single DRW and duplicating them. The two duplicated light curves are then multiplied by different magnifications computed assuming an untruncated Singular Isothermal Sphere (SIS) as the matter distribution for the lens, and sampling from a uniform distribution between 0 and 1 for the ratio between the source-lens angular separation and the lens Einstein radius. A delay sampled from a log-uniform distribution between a minute and two weeks is then introduced between the two light curves. The Bayes factor for all the simulated light curves is smaller than $B=1$, thus favouring the single DRW model with respect to the UBHD one, as expected.

Finally, in this work, we examined our proposed signature in single-band AGN light curves. However, the upcoming LSST will deliver multi-band light curves, which may be advantageous to combine. Therefore, we plan to extend the analysis presented in this paper to light-curves obtained in different wavelength bands \citep[as explored in][for the identification of periodicities in AGN light curves]{Covino2022}.

All of the above discussion and conclusions rest on the assumption that AGN intrinsic variability is well described by the DRW model. However, deviations from the DRW have been reported \citep[e.g.,][]{Zu2013, Kasliwal2015, Wang2019}, and alternative models have been proposed, such as the Damped Harmonic Oscillator (DHO) model \citep{Yu2022}. If AGN light curves follow a DHO process rather than a DRW, our results may be affected. In the overdamped DHO case \citep[see][]{Yu2022}, the covariance function reduces to a sum of two exponential kernels, resembling our UBHD model. Nevertheless, the two models predict different parameter correlations, which could, in principle, be used to distinguish between them. We plan to investigate the impact of including the DHO model on our results in future work.

\begin{acknowledgements}
      LB acknowledges ISCRA for awarding this project access to the LEONARDO supercomputer, owned by the EuroHPC Joint Undertaking, hosted by CINECA (Italy).
      
      LB wishes to thank the "Summer School for Astrostatistics in Crete" for providing training on the statistical methods adopted in this work. MC acknowledges support by the European Union (ERC, MMMonsters,
101117624).
FR acknowledges the support from the Next Generation EU funds within the National Recovery and Resilience Plan (PNRR), Mission 4 - Education and Research, Component 2 - From Research to Business (M4C2), Investment Line 3.1 - Strengthening and creation of Research Infrastructures, Project IR0000012 – “CTA+ - Cherenkov Telescope Array Plus.
MD acknowledge funding from MIUR under the grant
PRIN 2017-MB8AEZ, financial support from ICSC – Centro Nazionale di Ricerca in High Performance Computing, Big Data and Quantum Computing, funded by European Union – NextGenerationEU, and support by the Italian Ministry for Research and University (MUR) under Grant 'Progetto Dipartimenti di Eccellenza 2023-2027' (BiCoQ).
We acknowledge a financial contribution from the Bando Ricerca Fondamentale INAF 2022 Large Grant, {\textit{Dual and binary supermassive black holes in the multi-messenger era: from galaxy mergers to gravitational waves.}}
FR acknowledges support from ELSA. ELSA Euclid Legacy Science Advanced analysis tools” (Grant Agreement no. 101135203) is funded by the European Union. Views and opinions expressed are, however, those of the author(s) only and do not necessarily reflect those of the European Union or Innovate UK. Neither the European Union nor the granting authority can be held responsible for them. UK participation is funded through the UK Horizon guarantee scheme under Innovate UK grant 10093177.
\end{acknowledgements}

\bibliographystyle{aa} 
\bibliography{bibliography}

\begin{appendix}

\section{Periodogram computation} \label{app:periodogram_computation}
The periodogram is a useful tool to estimate the true PSD of a process \citep[see][]{Priestly1981, Bloomfield2000}.
In its classical definition, the periodogram can be defined as the square of the discrete Fourier transform \citep[see][]{Scargle1982}. Panels (b) and (d) of Figure \ref{fig:multiplot} show examples of the periodograms computed using the classical definition.

\begin{figure}[h!]
    \centering
    \includegraphics[width=\hsize]{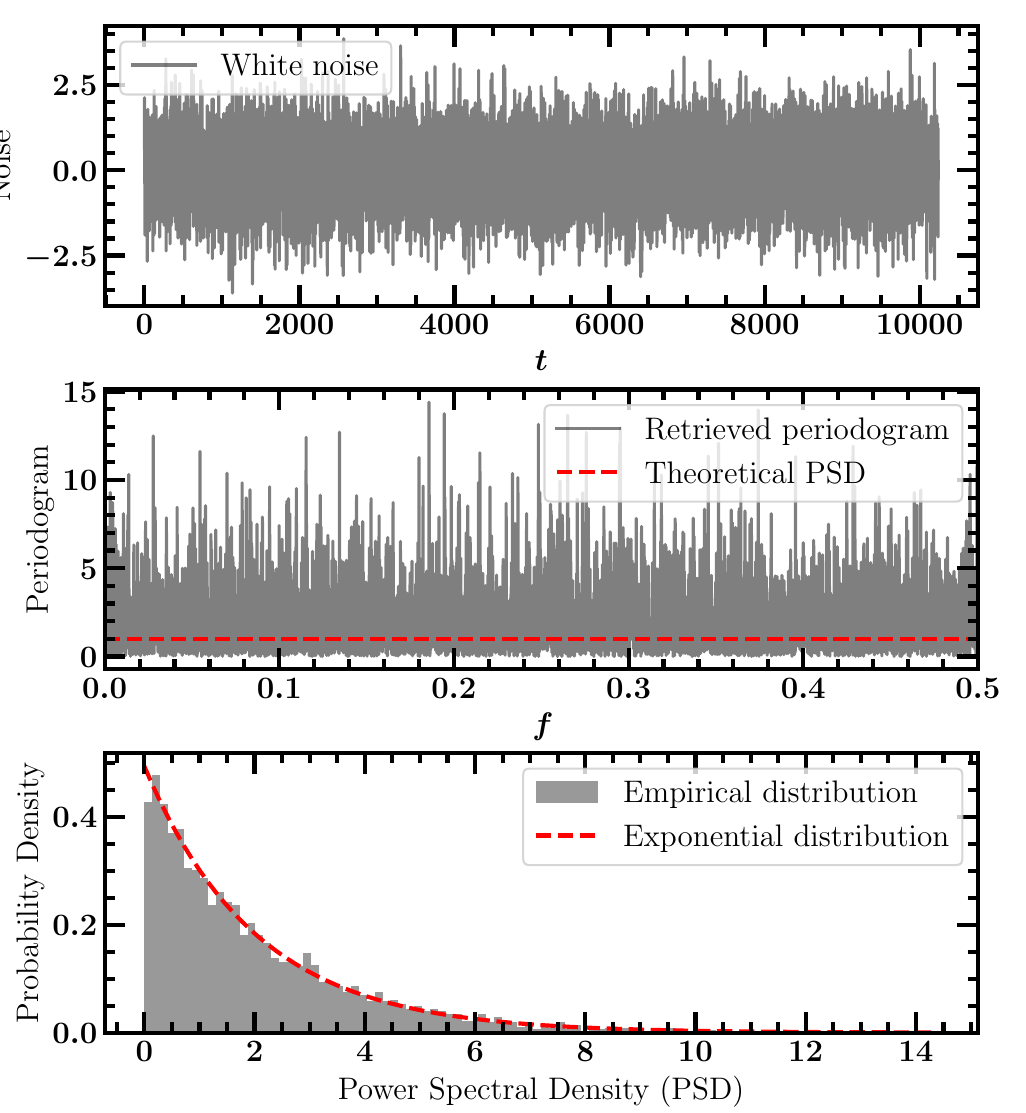}
    \caption{Example of a distribution for a white noise periodogram. Upper panel: noisy time series for a white noise with unity variance. Middle panel: comparison between the theoretical PSD (red dashed line) and the retrieved periodogram. Lower panel: distribution of the individual frequency bin estimates of the retrieved periodogram around the true PSD compared with the expected exponential distribution shown in Equation \ref{eq:exponential distribution} (red dashed lines).}
    \label{fig:white_noise}
\end{figure}

The sampled periodogram is not a smooth function of the frequencies but is scattered around the true PSD \citep[see][]{Vaughan2003}. More specifically, the periodogram computed at a given frequency is distributed as a $\chi^2$ distribution with two degrees of freedom, i.e. it is exponentially distributed as \citep[see][]{Scargle1982, vanderKlis1989}:
\begin{equation} \label{eq:exponential distribution}
    p(I_j/S_j)=\frac{1}{S_j}e^{-\frac{I_j}{S_j}}
\end{equation}
where $S_j$ and $I_j$ are respectively the true PSD and the value of the periodogram computed at the frequency $f_j=j/N \Delta t$ with $j=1...N/2$.
The shape of this distribution is due to the fact that the real and imaginary parts of the discrete Fourier transform are normally distributed for a stochastic process \citep[see][]{Jenkins1968, Priestly1981}. As an example, in Figure~\ref{fig:white_noise}, we show an example of a white noise light curve, its periodogram with the true PSD and the distribution of the periodogram around the true PSD, along with the theoretical $\chi^2$ distribution. 
The classical definition  
works particularly well for evenly sampled time series, but it is not without flaws. The periodogram is not a consistent estimator as the scatter does not decrease as the number of data points in the light curve increases \citep[][]{Jenkins1968}. Also, periodograms measured from finite data tend to be biased by windowing effects \citep[see][ and references therein]{Vaughan2003}. 

When the sampling is uneven, one can use the so-called Lomb-Scargle periodogram \citep[see][]{Lomb1976, Scargle1982}. The Lomb-Scargle periodogram is a generalisation of the classical periodogram, but its statistical properties are different. In particular, the observed periodogram is distorted from the underlying PSD by the sampling pattern of the underlying light curve. More specifically, in the context of red noise, effects such as aliasing and red noise leakage become important \citep[see][]{Papadakis95, Uttley2002}. Furthermore, the distribution of periodogram peaks is generally unknown, and correlations between the power at different frequency bins arise unless the light curve is evenly sampled and the periodogram is computed at the Fourier frequencies or another orthonormal frequency grid \citep[see][]{Vio_2010, Covino2022}.

\section{Fitting directly the PSD} \label{app:PSD_fitting}

\begin{figure*}[h!]
    \centering
    \begin{subfigure}[PSD fitting for a double DRW with similar variability amplitudes but considerably different damping timescales. This is the case in which the test can correctly identify the UBHD.]{
    \includegraphics[width=0.48 \hsize]{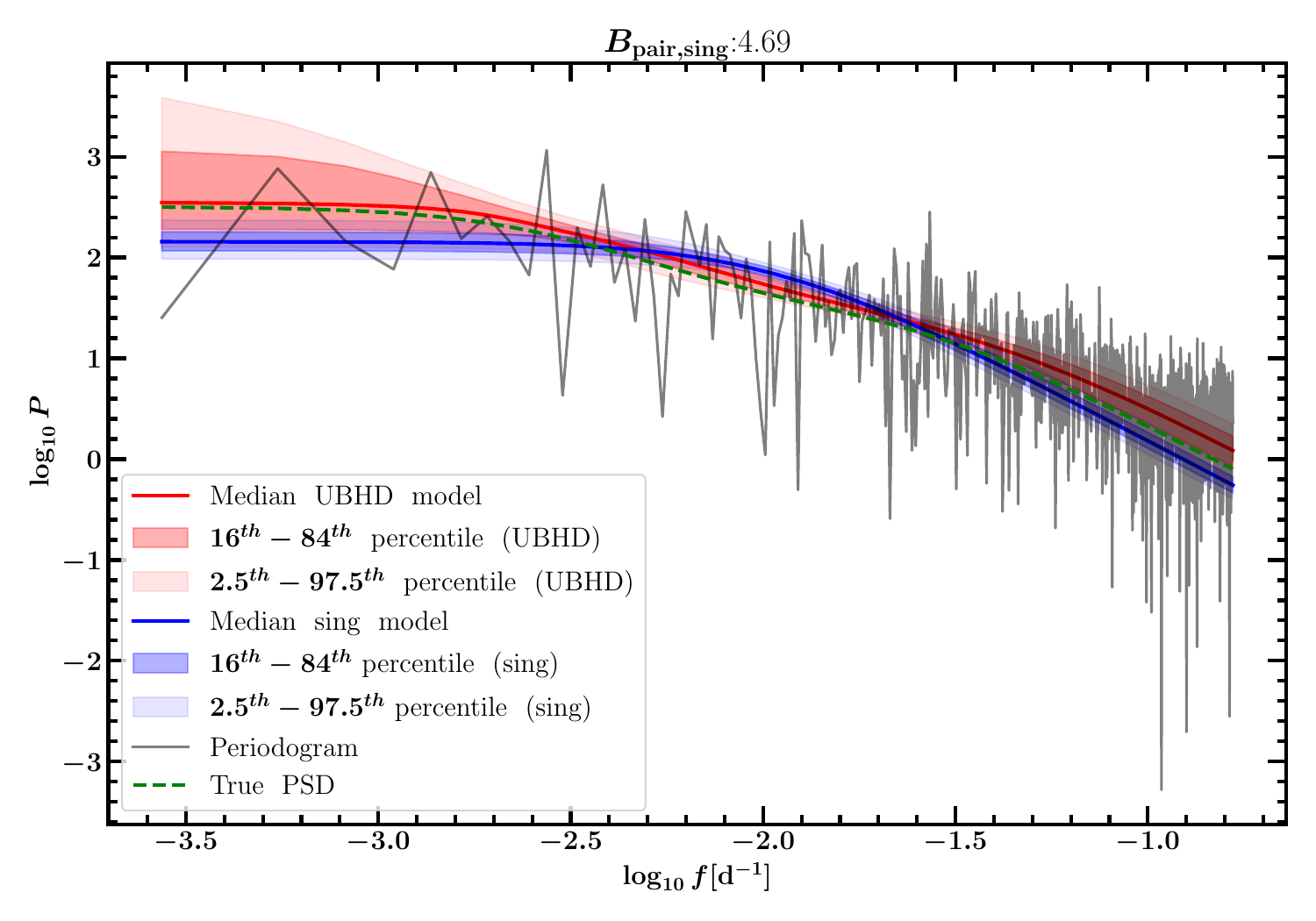}}
    \end{subfigure}
    \hfill
    \begin{subfigure}[PSD fitting for a double DRW with similar variability amplitudes and similar damping timescales.]{
    \includegraphics[width= 0.48 \hsize]{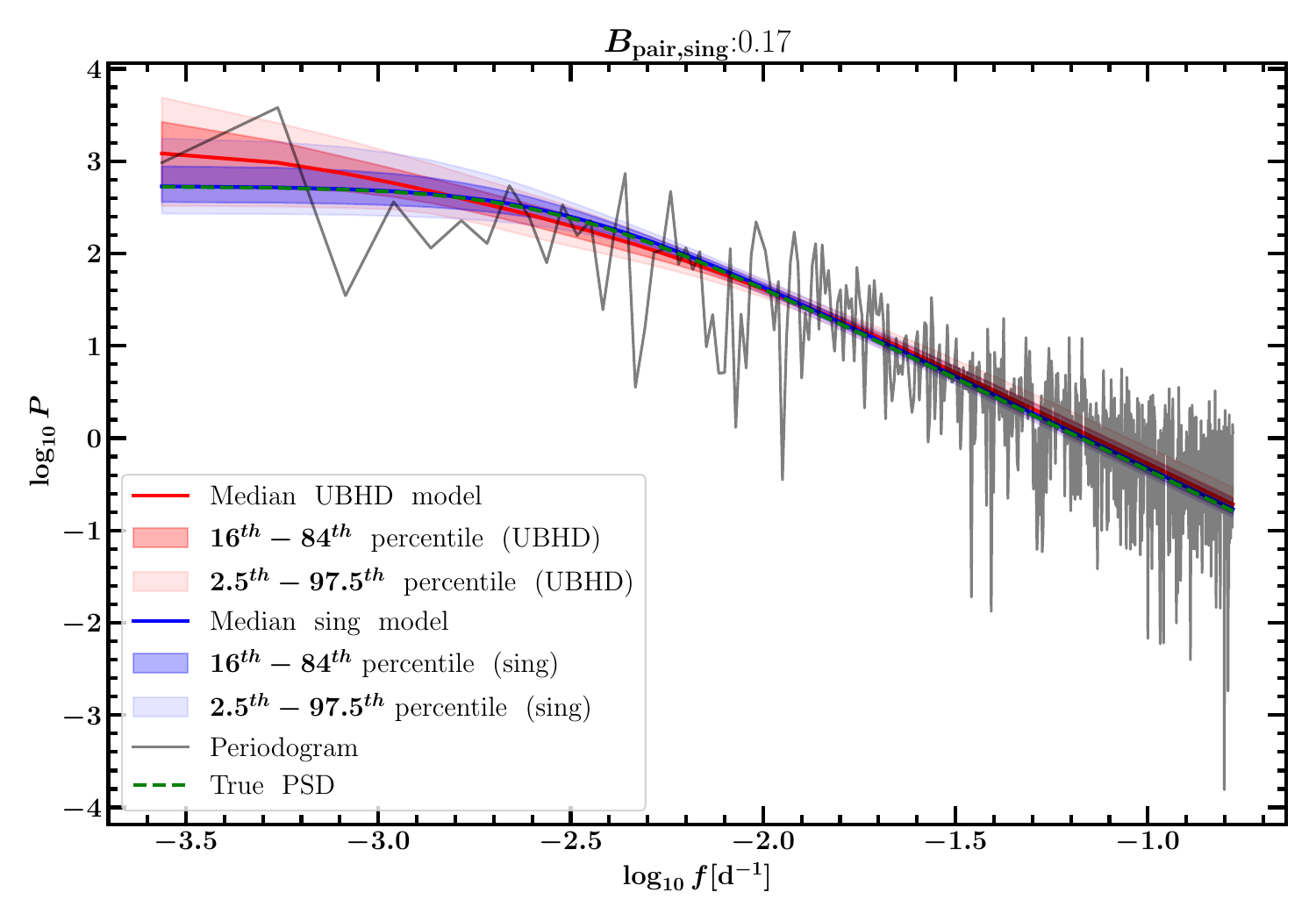}}
    \end{subfigure}
    \begin{subfigure}[PSD fitting for a double DRW  with considerably different variability amplitudes but similar damping timescales.]{
    \includegraphics[width= 0.48 \hsize]{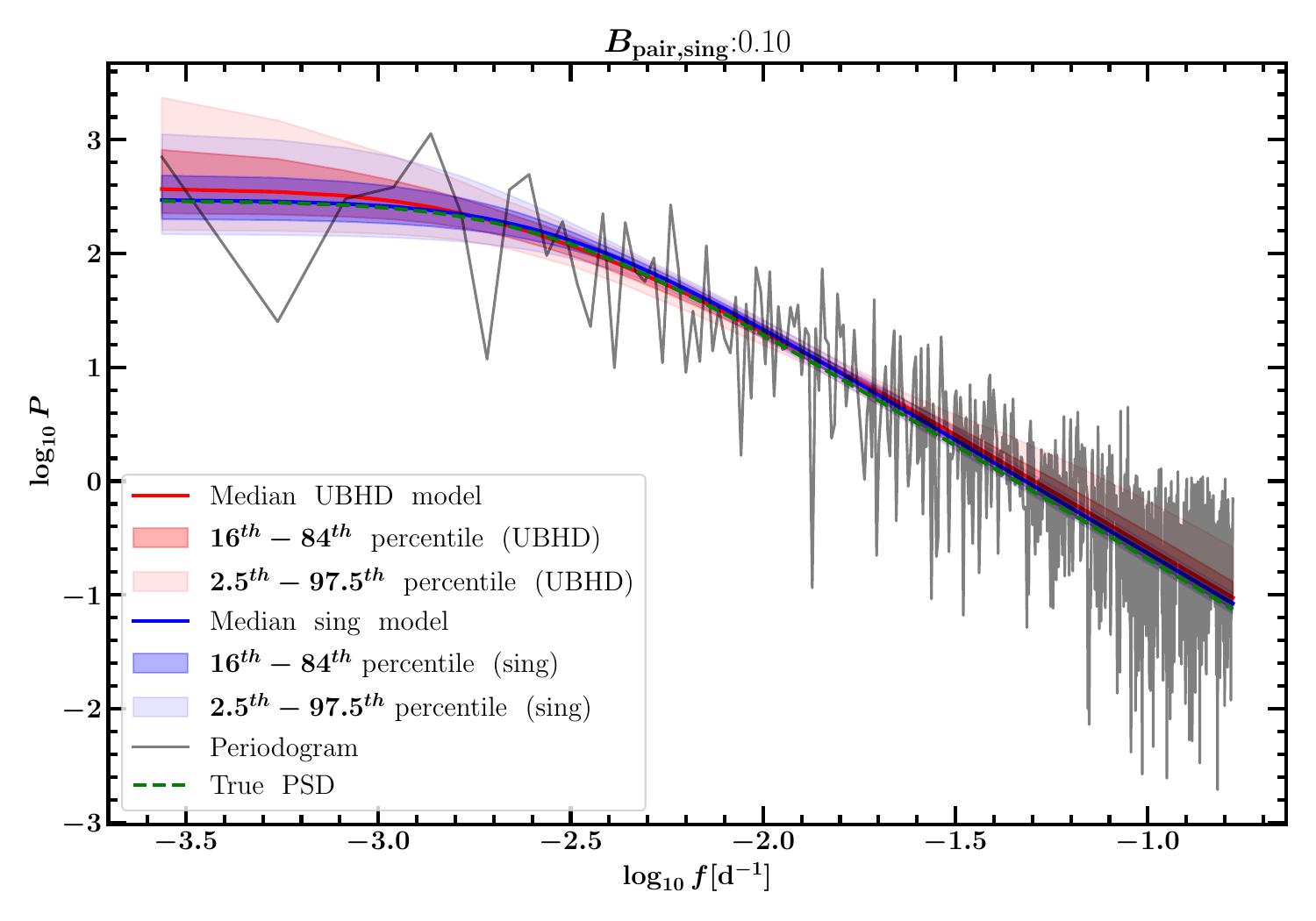}}
    \end{subfigure}
    \hfill
    \begin{subfigure}[PSD fitting for a double DRW kernel with considerably different variability amplitudes and damping timescales.]{
    \includegraphics[width=0.48\hsize]{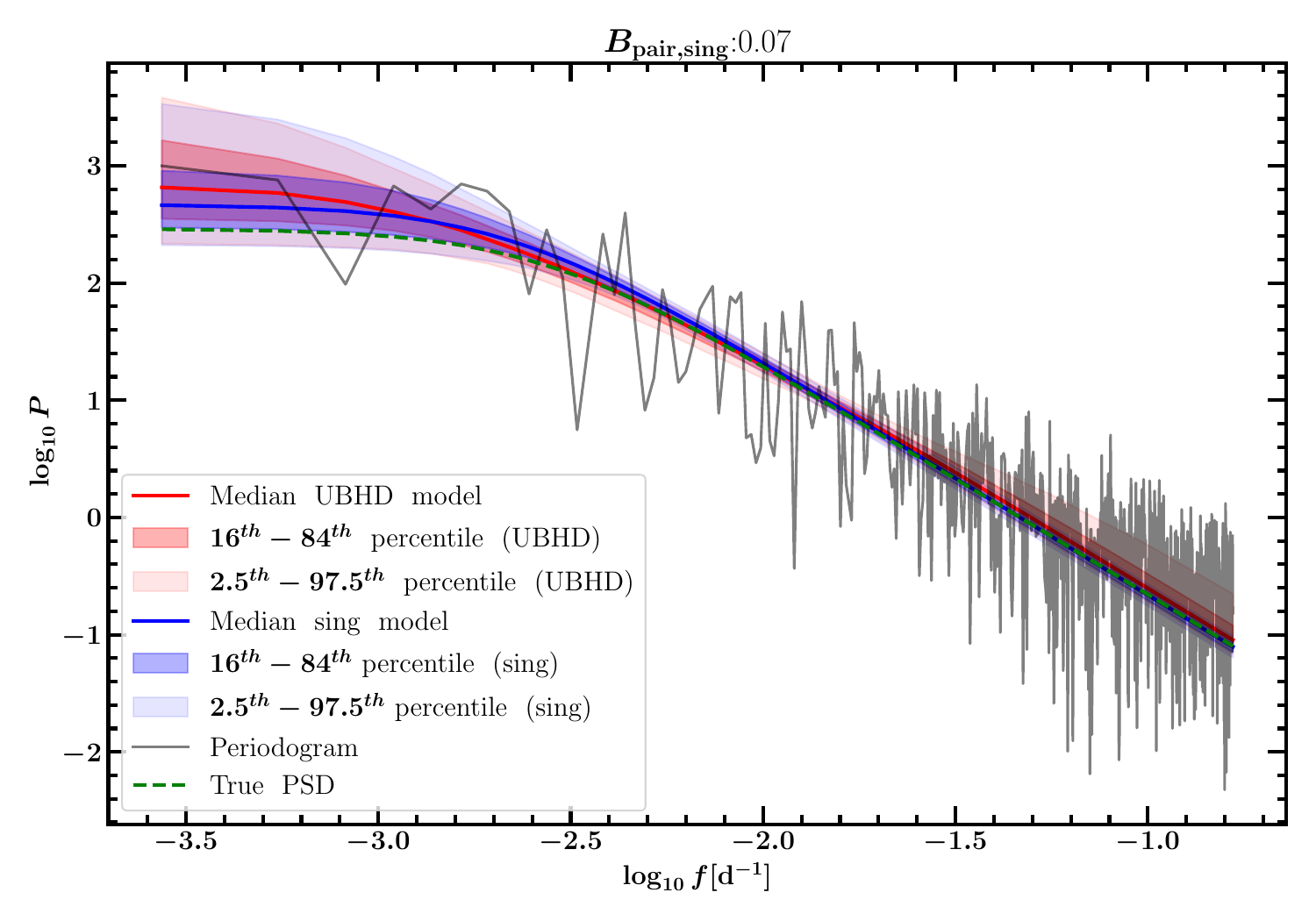}}
    \end{subfigure}
    \caption{Results of the direct fit of the PSD for evenly sampled light curves with a 3 d cadence for 10 yr. The grey line indicates the light curve periodogram. The red and blue solid lines show the median model for the double and single DRW, respectively. The red and blue shaded areas show confidence intervals ($16^{th}-84^{th}$ and $2.5^{th}-97.5^{th}$ ) for the UBHD and single DRW, respectively.}
    \label{fig:PSD_four_panel}
\end{figure*}

The best parameters of the stochastic process can be estimated directly from the periodogram. Also in this case, one can use nested sampling to do both evidence and best parameter estimation. In particular, the likelihood for the sampled periodogram points is given by the Whittle likelihood \citep[see][for an application to X-ray power spectra]{Barret2012}. Given the probability (\ref{eq:exponential distribution}), the Whittle log-likelihood can be written as:
\begin{equation}
    \log L=  \sum_{j=1}^{N-1} \frac{I_j}{S_j}+ \log S_j \ .
    \label{eq:Whittle_likelihood}
\end{equation}
This approach can be used only in light curves where the time is evenly sampled. 
In the case of uneven sampling, the computation of the periodogram through the Lomb-Scargle procedure introduces correlations between different frequencies so that the Whittle likelihood does not approximate the true likelihood of the periodogram anymore. In particular, the specific powers at each frequency bin are not always independent. In principle, the presence of a UBHD can be searched for by fitting a Lomb-Scargle periodogram from an unevenly sampled light curve, but the statistical issues mentioned above will prevent a robust statistical estimate of the evidence of the UBHD hypothesis. In the following, we present for completeness a few examples of evenly sampled mock UBHD light curves fitted with two red-noise PSDs, indicating the Bayes ratios obtained under the two assumptions\footnote{We stress again that the ratio obtained from fitting the Lomb-Scargle periodogram in unevenly sampled data should not be considered as a statistically robust indicator of the presence of a UBHD.}.
In Figure~\ref{fig:PSD_four_panel}, we show the result of the fitting procedure using the PSD and Whittle likelihood for a time series sampled every 3 d for 10 yr with the same parameter ratios used in Figure \ref{fig:double_kernel_PSD}. The dashed green line represents the true PSD, while the solid red and blue lines show the best fit from the UBHD and single MBH models, respectively. The red and blue shaded areas represent the $68\%$ and $95\%$ confidence intervals of the UBHD and single MBH models, respectively. Also in this case, we reported the Bayes ratio resulting from the nested sampling.

Another approach to assess the presence of a UBHD without a fully Bayesian analysis, which we do not explore here, is to generate a large number of simulated curves assuming a large grid of values for the parameters of the assumed noise model. By defining a suitable test-statistic, such as the $\chi^2$, one can find the set of parameters that best describe the observed light curve, and given a goodness-of-fit statistic, one can compare the results for different noise models. We note, though, that while such a procedure may reduce the computational cost of the fitting procedure, it typically provides only point estimates of the parameters and approximations to their uncertainties, and it does not fully account for correlations or degeneracies between parameters, in contrast to the fully Bayesian approach presented in this work.

\end{appendix}
\end{document}